\documentclass[11pt]{article}
\usepackage{amsmath}
\usepackage{amsfonts}
\usepackage{amssymb}
\usepackage{graphicx}

\newcommand{\Title}{\bf \sf \huge \noindent}
\newcommand{\Author}{\bf \sf \large \noindent}
\begin{document}

\mbox{  } \vskip 1.5cm

\Title{Do bosons obey Bose-Einstein distribution: two iterated
limits of Gentile distribution}

\noindent \Huge
------------------------------------------------------
\vskip 0.5cm
\Author{Wu-Sheng Dai and Mi Xie}

{\it \noindent \small Department of Physics, Tianjin University,
Tianjin 300072, P.
R. China}\\
{\it \noindent \small LiuHui Center for Applied Mathematics, Nankai
University \& Tianjin
University, Tianjin 300072, P. R. China}\\
{\it \noindent \small Email: \rm daiwusheng@tju.edu.cn,
xiemi@tju.edu.cn}

\thispagestyle{myheadings} \markboth{Preprint}{\underline{Physics
Letters A 373 (2009) 1524}} \vskip 1cm \rm \normalsize

\noindent {\bf Abstract:} It is a common impression that by only
setting the maximum occupation number to infinity, which is the
demand of the indistinguishability of bosons, one can achieve the
statistical distribution that bosons obey --- the Bose-Einstein
distribution. In this letter, however, we show that only with an
infinite maximum occupation number one cannot uniquely achieve the
Bose-Einstein distribution, since in the derivation of the
Bose-Einstein distribution, the problem of iterated limit is
encountered. For achieving the Bose-Einstein distribution, one needs
to take both the maximum occupation number and the total number of
particles to infinities, and, then, the problem of the order of
taking limits arises. Different orders of the limit operations will
lead to different statistical distributions. For achieving the
Bose-Einstein distribution, besides setting the maximum occupation
number, we also need to state the order of the limit operations.

\vskip 0.5cm \noindent  PACS codes: 05.30.Pr; 05.30.Jp

\vskip 0.5cm \noindent Keywords: iterated limit; Bose-Einstein
distribution; Gentile statistics; maximum occupation number
\newpage
\vskip 1cm \noindent
---------------------------------------------------------------------------------------------------------------------
\tableofcontents

\vskip 1cm \noindent
---------------------------------------------------------------------------------------------------------------------
\section{Introduction}

The indistinguishability of identical particles demands that for
Bose-Einstein statistics, a quantum state can be occupied by any
number of particles, and for Fermi-Dirac statistics, a quantum state
can be occupied by only one particle. Starting from this point, as a
common impression, in statistical mechanics, one can obtain
statistical distributions by only setting the various values of the
maximum occupation number $n$: by $n\rightarrow\infty$, one achieves
the Bose-Einstein distribution, and by $n=1 $, one achieves the
Fermi-Dirac distribution. In this letter, however, we show that,
contrary to this common impression, only from a maximum occupation
number, one cannot uniquely construct the statistical distribution
in the Bose-Einstein case, though such a treatment works well in the
Fermi-Dirac case. This difference between Bose-Einstein and
Fermi-Dirac cases comes from the fact that the maximum occupation
number in the Bose-Einstein case is $\infty$, but in the Fermi-Dirac
case is a finite number.

In section \ref{order}, we point out that there exists not only one
possibility to take limits in the derivation of the Bose-Einstein
distribution. In sections \ref{BEd} and \ref{bBEd}, we discuss the
various orders of taking limits and show which one will lead to the
Bose-Einstein distribution. The conclusion is summarized in Sec.
\ref{Conclusions}.

\section{The order of taking limits\label{order}}

The reason why the Bose-Einstein distribution cannot be uniquely
achieved by only setting $n\rightarrow\infty$ is that there are two
infinities involved in the derivation of the Bose-Einstein
distribution --- the maximum occupation number and the total number
of particles. The infinite maximum occupation number is the demand
of the indistinguishability of bosons; the infinite total number of
particles, i.e., the thermodynamic limit, guarantees the extensivity
of a thermodynamic system. As a result, during the derivation, we
need to perform two limits, $\lim\limits_{n\rightarrow\infty}$ and
$\lim \limits_{\left\langle N\right\rangle \rightarrow\infty}$,
where $\left\langle N\right\rangle $ denotes the mean total number
of particles in the grand canonical ensemble. This means that in the
Bose-Einstein case, we have to first make a choice of the order of
the limit operations.

For simultaneously taking the maximum occupation number $n$ and the
total number of particles $\left\langle N\right\rangle $ into
account, we start with Gentile statistics \cite{Gentile,Khare} in
which the maximum occupation number $n$ can take on any value.
Different values of $n$ correspond to different kinds of statistics,
and the Bose-Einstein case is a special case of Gentile statistics
with $n\rightarrow\infty$ \cite{OursAnn}. The Gentile distribution
reads%
\begin{equation}
f_{G}=\frac{1}{z^{-1}e^{\varepsilon/\left(  kT\right)
}-1}-\frac{n+1}{\left[
z^{-1}e^{\varepsilon/\left(  kT\right)  }\right]  ^{n+1}-1},\label{fGp}%
\end{equation}
where $T$ is the temperature, $\varepsilon$ is the energy, and $z$
is the fugacity. In the grand canonical ensemble, the fugacity $z$
is determined by \cite{OursAnn}
\begin{equation}
\left\langle N\right\rangle =\frac{V}{\lambda^{3}}h_{3/2}\left(
z\right)
+\frac{1}{z^{-1}-1}-\frac{n+1}{z^{-\left(  n+1\right)  }-1},\label{meanN}%
\end{equation}
where $\lambda=h/\sqrt{2\pi mkT}$ is the mean thermal wavelength,%
\[
h_{3/2}\left(  z\right)  =g_{3/2}\left(  z\right)  -\frac{1}{\sqrt{n+1}%
}g_{3/2}\left(  z^{n+1}\right)  ,
\]
and $g_{3/2}\left(  z\right)  $ is the Bose-Einstein integral. In
principle,
from eq. (\ref{meanN}), one can solve the expression of $z$ as%
\begin{equation}
z=w\left(  T,n,\left\langle N\right\rangle \right)  ,
\end{equation}
and then write the distribution function as
\begin{align}
f_{G}\left(  T,n,\left\langle N\right\rangle \right)   &
=\frac{1}{w\left( T,n,\left\langle N\right\rangle \right)
^{-1}e^{\varepsilon/\left(
kT\right)  }-1}\nonumber\\
&  -\frac{n+1}{\left[  w\left(  T,n,\left\langle N\right\rangle
\right)
^{-1}e^{\varepsilon/\left(  kT\right)  }\right]  ^{n+1}-1}.\label{fG}%
\end{align}

There are two infinities, $n$ and $\left\langle N\right\rangle $, in
the distribution function $f_{G}\left(  T,n,\left\langle
N\right\rangle \right) $. For recovering the Bose-Einstein
distribution, we need to take both $n$ and $\left\langle
N\right\rangle $ tend to $\infty$ in eq. (\ref{fG}). However,
there are three possible ways to take the limits: two iterated limits%
\begin{equation}
\lim\limits_{\left\langle N\right\rangle \rightarrow\infty}\lim
\limits_{n\rightarrow\infty}f_{G}\left(  T,n,\left\langle
N\right\rangle
\right) \label{BE}%
\end{equation}
and%
\begin{equation}
\lim\limits_{n\rightarrow\infty}\lim\limits_{\left\langle
N\right\rangle \rightarrow\infty}f_{G}\left(  T,n,\left\langle
N\right\rangle \right)
,\label{BEb}%
\end{equation}
and the double limit%
\begin{equation}
\underset{\left\langle N\right\rangle \rightarrow\infty}{\lim
\limits_{n\rightarrow\infty}}f_{G}\left(  T,n,\left\langle
N\right\rangle
\right)  .\label{BEdouble}%
\end{equation}
The first thing is thus to choose the way how to take the limit from
the above three. Nevertheless, the indistinguishability of bosons
only requires an infinite maximum occupation number, but does not
say anything about the order of the limits. Generally speaking,
different orders of the two limit operations will lead to different
results \cite{Knapp,Jaynes}. At first sight, one may think that a
natural choice is the double limit (\ref{BEdouble}). However, in the
following, we will show that the double limit (\ref{BEdouble}) does
not exist, since the iterated limits (\ref{BE}) and (\ref{BEb})
exist but are not equal to each other. It is the iterated limit
(\ref{BE}) that leads to the Bose-Einstein distribution. That is to
say, for obtaining the Bose-Einstein distribution, besides setting
$n\rightarrow\infty$, the demand of indistinguishability, we also
need an additional condition that the order of the iterated limit
should be $\lim\limits_{\left\langle N\right\rangle
\rightarrow\infty}\lim\limits_{n\rightarrow\infty}$.

\section{$\lim\limits_{\left\langle N\right\rangle \rightarrow\infty}%
\lim\limits_{n\rightarrow\infty}f_{G}$: the Bose-Einstein
distribution\label{BEd}}

We first discuss the case of the iterated limit
$\lim\limits_{\left\langle N\right\rangle
\rightarrow\infty}\lim\limits_{n\rightarrow\infty}f_{G}$. In this
case, the limit $\lim\limits_{n\rightarrow\infty}$ is first
performed.

In principle, for obtaining the statistical distribution under the
limit of $n\rightarrow\infty$, we need to calculate the explicit
expression for the fugacity $z$, substitute it into eq. (\ref{fGp})
to achieve eq. (\ref{fG}), and then perform the limit
$\lim\limits_{n\rightarrow\infty}$ in eq. (\ref{fG}). However, it is
difficult to obtain the expression of $z$, so we will do this in an
indirect way.

We first determine the maximum value of the fugacity $z$ at
$n\rightarrow \infty$. Since
$\frac{1}{z^{-1}-1}-\frac{n+1}{z^{-\left(  n+1\right)  }-1}$ is a
monotonically increasing function of $z$, from eq. (\ref{meanN}),
the
maximum value of $z$ appears at $T=0$, or, $\lambda\rightarrow\infty$, i.e.,%
\begin{equation}
\left\langle N\right\rangle =\lim\limits_{n\rightarrow\infty}\left[
\frac {1}{z_{\max}^{-1}-1}-\frac{n+1}{z_{\max}^{-\left(  n+1\right)
}-1}\right]
.\label{zmax}%
\end{equation}
The solution of eq. (\ref{zmax}) is%
\begin{equation}
z_{\max}=\frac{\left\langle N\right\rangle }{\left\langle
N\right\rangle +1}.
\end{equation}
Based on this result, we can take the limit
$\lim\limits_{n\rightarrow\infty}
$ on eq. (\ref{meanN}):%
\begin{equation}
\left\langle N\right\rangle =\frac{V}{\lambda^{3}}g_{3/2}\left(
z\right)
+\frac{1}{z^{-1}-1},\label{BEN}%
\end{equation}
or%
\begin{equation}
1=\frac{1}{\rho\lambda^{3}}g_{3/2}\left(  z\right)
+\frac{1}{\left\langle
N\right\rangle }\frac{1}{z^{-1}-1},\label{ninf}%
\end{equation}
where $\rho=\left\langle N\right\rangle /V$ is the particle number
density.

Now, we can take the limit $\lim\limits_{\left\langle N\right\rangle
\rightarrow\infty}$ and obtain the range of value of the fugacity%
\begin{equation}
0<z\leq1.
\end{equation}
This is just the Bose-Einstein case \cite{Huang}. The corresponding
statistical distribution is the Bose-Einstein distribution:%
\begin{equation}
f=\frac{1}{z^{-1}e^{\varepsilon/\left(  kT\right)  }-1}.
\end{equation}

The above result shows that if the order of the limits is chosen as
$\lim\limits_{\left\langle N\right\rangle \rightarrow\infty}\lim
\limits_{n\rightarrow\infty}f_{G}$, one achieves the Bose-Einstein
distribution.

\section{$\lim\limits_{n\rightarrow\infty}\lim\limits_{\left\langle
N\right\rangle \rightarrow\infty}f_{G}$: beyond the Bose-Einstein
distribution\label{bBEd}}

An alternative choice for the iterated limit is
$\lim\limits_{n\rightarrow \infty}\lim\limits_{\left\langle
N\right\rangle \rightarrow\infty}$, i.e., the limit
$\lim\limits_{\left\langle N\right\rangle \rightarrow\infty}$ is
first performed.

Rewrite eq. (\ref{meanN}) as%
\begin{equation}
1=\frac{1}{\rho\lambda^{3}}h_{3/2}\left(  z\right)
+\frac{1}{\left\langle N\right\rangle }\left[
\frac{1}{z^{-1}-1}-\frac{n+1}{z^{-\left(  n+1\right)
}-1}\right]  .\label{meanNdf}%
\end{equation}
Here $\frac{1}{z^{-1}-1}-\frac{n+1}{z^{-\left(  n+1\right)  }-1}$ is
the number of the particles occupying the ground state, whose
maximum value is the maximum occupation number $n$. When taking the
limit $\lim \limits_{\left\langle N\right\rangle
\rightarrow\infty}$, the ground-state term vanishes. Consequently,
taking the limit $\lim\limits_{\left\langle N\right\rangle
\rightarrow\infty}$ on eq. (\ref{meanNdf}) gives
\begin{equation}
\rho\lambda^{3}=h_{3/2}\left(  z\right)  .\label{Ninf}%
\end{equation}
$h_{3/2}\left(  z\right)  $ is a monotonically increasing function
of $z$ and $0<h_{3/2}\left(  z\right)  <\infty$ \cite{OursAnn}, so
the range of the value of $z$ is $0<z<\infty$.

Next, take the limit $\lim\limits_{n\rightarrow\infty}$ on eq.
(\ref{Ninf}).

After taking the limit $\lim\limits_{\left\langle N\right\rangle
\rightarrow\infty}$, eq. (\ref{meanNdf}) becomes eq. (\ref{Ninf}).
The\ left-hand side of eq. (\ref{Ninf}) can take on any value from
$0$ to $\infty$, so the fugacity $z$ can take on the value that is
greater than $1$ due to the fact that $h_{3/2}\left(  z\right)  $ is
a monotonically increasing
function of $z$ and when $z=1$, $\lim\limits_{n\rightarrow\infty}%
h_{3/2}\left(  z\right)  =\zeta\left(  3/2\right)  $, where
$\zeta\left( 3/2\right)  $ is the zeta function.

The distribution corresponding to $z>1$ must not be the
Bose-Einstein distribution because in the Bose-Einstein case,
$0<z\leq1$. Concretely, taking $\lim\limits_{n\rightarrow\infty}$ in
the distribution function gives
\begin{equation}
f_{G}=\left\{
\begin{array}
[c]{ll}%
\frac{1}{z^{-1}e^{\varepsilon/\left(  kT\right)  }-1}, & \varepsilon
>\mu\left(  T\right)  ,\\
\infty, & \varepsilon<\mu\left(  T\right)  ,
\end{array}
\right. \label{fbeyond}%
\end{equation}
where $\mu$ is the chemical potential. Notice that before taking
$\lim\limits_{n\rightarrow\infty}$, the limit
$\lim\limits_{\left\langle N\right\rangle \rightarrow\infty}$ has
been already taken. Eq. (\ref{fbeyond}) is a statistical
distribution with an infinite maximum occupation number, i.e.,
$n\rightarrow\infty$, but is not the Bose-Einstein distribution.

The above results show that both the iterated limits $\lim
\limits_{\left\langle N\right\rangle \rightarrow\infty}\lim
\limits_{n\rightarrow\infty}f_{G}$ and $\lim\limits_{n\rightarrow\infty}%
\lim\limits_{\left\langle N\right\rangle \rightarrow\infty}f_{G}$
exist, but are not equal to each other. As a result, the double
limit $\underset {\left\langle N\right\rangle
\rightarrow\infty}{\lim\limits_{n\rightarrow \infty}}f_{G}$ does not
exist.

\section{Conclusions\label{Conclusions}}

In conclusion, in the derivation of the Bose-Einstein distribution,
two infinities are encountered. Thus, we have to choose the way how
to take the limits. Different orders of the limit operations give
different statistical distributions. In the common derivation of
Bose-Einstein distribution, the order of the iterated limit has been
unstatedly chosen as $\lim \limits_{\left\langle N\right\rangle
\rightarrow\infty}\lim \limits_{n\rightarrow\infty}$. That is to
say, to achieve the Bose-Einstein distribution, besides setting the
maximum occupation number to infinity, one also needs to choose the
order of taking limits.\vspace{0.2in}

\textbf{Acknowledgements} We are very indebted to Dr. G. Zeitrauman
for his encouragement. This work is supported in part by NSF of
China under Grant No. 10605013 and the Hi-Tech Research and
Development Programme of China under Grant No. 2006AA03Z407.

\end{document}